\documentstyle[12pt,aaspp]{article}

\input{epsf}
\newcommand{\lapprox}{\raisebox{-.5ex}{ $\stackrel{<}{\sim}$\/}}
\newcommand{\gapprox}{\raisebox{-.5ex}{ $\stackrel{>}{\sim}$\/}}
\begin{document}

\title{The Discovery of Halley-sized Kuiper Belt Objects
Using HST}
\author{Anita L. Cochran\altaffilmark{1}}
\affil{Astronomy Department \& McDonald Observatory, University
of Texas, Austin, TX 78712}
\author{Harold F. Levison\altaffilmark{1} and S. Alan Stern\altaffilmark{1}}
\affil{Geophysical, Astrophysical, \& Planetary Sciences, Southwest Research
Institute,\break 1050 Walnut St, Suite 429, Boulder, CO  80302}
\author{Martin J. Duncan\altaffilmark{1}}
\affil{Physics Department, Queens University, Kingston, Ontario  K7L~3N6}

\altaffiltext{1}{Based on observations with NASA/ESA Hubble Space
Telescope obtained at the Space Telescope Science Institute, which is
operated by Universities for Research in Astronomy, Incorporated, under
NASA contract NAS5-26555.}

\begin{abstract} We report the statistical detection (at the $>99$\%
confidence level) of a population of 28th magnitude objects exhibiting
proper motions of $\approx$1 arcsecond per hour at quadrature in deep
HST/WFPC2 images.  The drift directions imply a preponderance of
objects on prograde orbits concentrated near the ecliptic plane.  We
interpret this as the detection of objects which reside in the
Kuiper belt near or beyond the orbit of Neptune, comparable in size to
comet 1P/Halley (radii $\sim 10$\,{\sc km} for an albedo of 4\%).  Our
observations imply a population of $\sim 2.5\times10^4$ objects~deg$^{-2}$
with $V<28.6$.  This must be viewed as a lower limit until our search
of parameter space is complete.  Our observations imply that there are
$>2\times 10^8$ objects of this size in the Kuiper belt with
inclinations $<12^\circ$ and within $\sim 40$\,{\sc au} of the Sun.

\end{abstract}


\section{Introduction}

Recently, over 20 objects with radii of 100--200\,{\sc km} have been
discovered beyond the orbit of Neptune that may represent the
brightest members of a heretofore undetected class of objects in the
solar system (Jewitt \& Luu~1995; also Stern~1995, Weissman~1995).
Such a population has been suspected since
Edgeworth~(1949) and Kuiper~(1951) independently pointed out that it
seems unlikely that the disk of planetesimals that formed the planets
would have abruptly ended at the edge of the planetary system, and
that originally there was a significant number of
planetesimals ($i.e.$ comets) in near-circular orbits outside the planetary
region.

Fernandez (1980) suggested that such a belt of distant icy
planetesimals could serve as the source of the short-period comets and
would be more dynamically efficient than evolving long-period comets
inward from the Oort cloud as first suggested by Newton (1893; also
Everhart 1972, 1977).  Duncan {\it et al.} (1988) confirmed this with
dynamical simulations which showed that a cometary source with a low
initial inclination distribution was far more consistent with the
observed orbits of the short-period comets than the randomly
distributed inclinations of comets in the Oort cloud.  Subsequent
numerical integrations have shown that a significant fraction of the
objects that formed in the Kuiper belt are stable for the age of the
solar system but that weak gravitational instabilities provide a large
enough influx to replenish the current population of short-period comets if
the population of the Kuiper belt region is $\sim10^{10}$ objects
(Levison \& Duncan~1993, Holman \& Wisdom~1993, Duncan {\it et al.}
1995).  However, the detection of objects the size of short-period
comets in the Kuiper belt, which is necessary to confirm these
predictions, has been beyond observational reach until very recently.
The theoretical work referred to above did, however, spur searches for
larger objects.  Such objects were first discovered in 1992 (Jewitt \&
Luu~1992).  As of this writing, 24 such objects have been discovered.
Of these, 11 have been observed over a long enough period of time to
have fairly well determined orbits (Marsden, priv.~comm.). All 11 are
members of the Kuiper belt.  Comparisons with numerical integrations
(Duncan {\it et al.} 1995) show that their orbits are dynamically
stable for the age of the solar system.  The discovery of these
100--200\,{\sc km}-sized objects proved the Kuiper belt is populated.
It did not, however, provide the observational link to show that the
Kuiper belt is the major source of the short-period comets because
comets are so much smaller. Here we present HST/WFPC2 observations
showing there is a large population of Halley-sized objects (radii
$\sim\,10$\,km) in this region.

\section{Observations and Reductions}

On 21.7--23.0 August 1994 UT, we obtained thirty-four exposures of a
field on the ecliptic near morning quadrature with the ``Wide V''
filter on HST/WFPC2.  Each exposure was 500--600\,sec, for a total
integration time of 5\,hours.  The observations were of a single field
of 4 arcmin$^2$ centered at R.A.= 3 41 20, Dec = +19 34.7 (equinox
2000.0).  This field was chosen because it was along the ecliptic and
prior observations had shown that it contained very few stars and galaxies,
which helped minimize confusion from background
objects.  Thirty hours elapsed from the start of the first exposure
to the end of the last.  For an outer solar system body observed at
quadrature, the Earth's parallactic motion is near zero so that any
apparent motion of the body is attributable to the body's orbit.  For a
body in the Kuiper belt the orbital motion is $\lapprox 1$\,arcsec
hr$^{-1}$.  With the plate scale of the WF chips, this
orbital motion is equal to $\lapprox 10$ pixels hr$^{-1}$, $\sim 1$
pixel in each ten minute exposure, or a total of up to 300 pixels in 30
hours.

In principle, one could inspect a very deep exposure to
identify faint trails in the Kuiper belt.  However, each WFPC2 frame is
littered with hundreds of radiation events.  Some appear as
single pixels or clusters of pixels; others appear as streaks. It was
necessary to remove these prior to searching for Kuiper belt objects.
Standard techniques for removing radiation events
employ a median sum of many images, and rely on
the concept that real objects remain fixed relative to each other, but
radiation events appear in random pixels.  Unfortunately, the objects
we are searching for move relative to background stars and galaxies,
and would be removed by these standard methods.  In addition to these
difficulties, we also
needed to remove from our images any objects that do {\it not} move in
the fields as a function of time.  To solve both these problems we
adopted the following strategy.  We produced a median sum of all 34
exposures.  This left a high signal/noise image with all of the stars
and galaxies but with no Kuiper belt objects and no radiation
signatures.  Then, this median sum image was normalized and subtracted
from each of the individual exposures.  At this point, we had 34 images
with just radiation events, moving objects, and noise.

The next step was to combine the images so that only the Kuiper belt
objects remained.  First, we shifted the images so that a Kuiper belt
object appeared in the same pixel in all of the frames.  To accomplish
this, we specified various trial orbits as described
in detail in the next section.  For any orbit, one can
compute the drift rate of an object moving on that orbit during the
interval of our observations.
We shifted each image to compensate for the drift rate to
produce 34 images which were all co-aligned to one of the images for
the desired orbit.  Then we combined the shifted images into a median
sum so that a Kuiper belt object with the predicted drift rate would be
in the same pixel for each shifted image, and thus would remain in the
sum.  In contrast, random radiation events would be unlikely to
co-align.  Indeed, if a particular physical pixel of the CCD were
always high, that pixel would not create a detectable signal in the
median sum since it would be shifted into different places.
This process also removed main belt asteroids which traversed our
field since they appear as streaks.

After computing the image medians, we found by inspection that there
were no obvious bright objects in our field.  Therefore, we developed
an automatic search routine to examine each of the summed images for faint
objects.  For each pixel in the image, we computed the average signal
for a five pixel pattern (a ``plus'') centered on the pixel.  This was
compared with the local background which was computed as the mean of a
$15\times15$ pixel box centered on the pixel under study.  If the plus
average was higher than at least 1.5$\sigma$ of the background, this
pixel was flagged for further investigation.  Our results are not
sensitive to the choice of the 1.5$\sigma$ limit since our
ultimate detection limit was set by our further processing and calibration.

We needed to develop a mechanism for determining which of the resultant
``objects'' were real and which were just coincidental alignments of
noise.  This goal was complicated by the fact that the PSF of a star in
our frames was only $1.3\,$ pixels FWHM, so that
it is not possible to distinguish definitely between noise spikes and
real objects by the width of their images.  Thus, we adopted the
following strategy.  From the 34 individual exposures
we created 6 images, each consisting of half of the exposures.
The first of the six contains the first seventeen exposures.  The
second contains the last seventeen.  The third and fourth
contain the even and odd exposures, respectively.  The fifth and sixth
images contain $1,2,3,\ 7,8,9,\ \dots$ and $4,5,6,\ 10,11,12,\ \dots$,
respectively.  We require that something show up in the same pixel
in at least four of these six images for it to be retained for further
statistical analysis.  As with our choice of $\sigma$ above, the
condition that a candidate be in at least four (as opposed to three or
five) images did not significantly affect our results.  Something in
fewer than four images would have been rejected later by the
statistical analysis.

In order to understand our detection threshold, we ran tests with
artificial data.  Using the calibration supplied by STScI, we
implanted objects of known magnitudes from V=23 to 29.5 into the original
34 exposures.  Each object was implanted at a different pixel in each exposure
to simulate it moving on a Kuiper belt orbit.  We then processed these
images using our standard reduction procedure as described above. We found
that the automatic search was complete to V=27.7.  By V=28.6, this
search detected only 50\% of the artificial objects.  No artificial
objects fainter than V=29.5 were detected.  To be conservative, we
defined the ``limiting magnitude'' of our search as V=28.6, which
corresponds to a detection probability of 50\%.  (Hainaut {\it et al.}
(1994) ran a similar test on their NTT data and concluded that their
detection threshold was at the 90\% probability level.  Since they were
searching for specific objects rather than comparing with a control
sample, they could not allow for false detections in their analysis;
our statistical approach does not prohibit them.)  With these
artificial data, we were thereby able to calibrate the magnitudes of
the real objects which we detected.

When searching for objects so close to the detection threshold, it is
possible that coincidental alignments of noise will occasionally
occur.  To quantify
how many of our ``detections'' were likely to be false
alarms, we proceeded as follows.  To understand the noise properties,
we shifted and co-added images using orbits which were non-viable
Kuiper belt orbits, such as retrograde orbits.  We do not expect such
orbits to be populated because the Kuiper belt is a disk.
For every Kuiper belt orbit for which we produced the
six sums described above, we also produced the comparable six sums for
the same orbit but with the velocity of the object reversed.  These
were then searched with our automatic search routine in the same manner
as the true orbit images.  The results of these searches were analyzed
in the same statistical manner as the true orbits.  There was no
difference in the processing of the prograde and retrograde orbits
other than the sign of the velocity, thus nothing which we did should
produce systematic differences.

\section{Results and Analysis}

Because an object in the Kuiper belt would orbit the Sun once every $\sim250$
years but our observations only span 30 hours,  we cannot
uniquely constrain the orbit of any particular object.  However, our
reduction process produces a distribution of objects as a
function of various orbital parameters.
As noted above, we must specify the orbit of the objects that we are
searching for
before we begin our reduction procedures.  The ability to detect an
object is strongly sensitive to our choice of the assumed orbit, and
thus a complete search of the images requires a very fine grid in
orbital element space.
The only ``observables'' of the orbit are its drift rate with
respect to the background stars ($\dot\theta$) and the angle the
orbit makes with the ecliptic ($\phi$).  We found that in order to
find all the objects and yet not find an object more than once, a grid
with a resolution of $\sim 0.1$\,arcsec hr$^{-1}$ in $\dot\theta$ and
$\sim 1^\circ$ in $\phi$ is required.  In the long run, we plan to
study $\sim 500$ possible prograde orbits.  So far, we have studied
154 prograde orbits and their
154 retrograde counterparts.

We present the results of a search for objects with $ \dot\theta
\in [0.73,1.16]$\,arcsec hr$^{-1}$ and $\phi \in [0,30.3]$ deg.  We
chose this $\dot\theta$ range because it is the drift rate of objects
at perihelion in Neptune's 2:3 mean motion resonance ($a=39.4$\,{\sc
au}) for orbital eccentricities of 0.1 to 0.3.  We interpret the drift
rates in terms of objects in this resonance for two reasons.  First,
the 2:3 mean motion resonance with Neptune is known to be populated.
(Pluto is in this resonance, as are at least 2 and perhaps as many as 8
of the 24 known 100-200\,{\sc km}-sized Kuiper belt objects; Marsden
1994, 1995).  Second, objects in these orbits have the smallest
perihelion distances in the stable Kuiper belt (Duncan {\it et al.}
1995, Morbidelli {\it et al.} 1995) and thus, an object of a given size
appears brightest in these orbits. We note,
however, that the drift rates we studied are consistent with many other
orbits which are not in a 2:3 mean motion resonance with Neptune.  For
example, our range in $\dot\theta$ is consistent with circular orbits
with semi-major axis, $\sim 25<a<32$ {\sc au}, and
with objects at perihelia in parabolic orbits with perihelion distances
between $\sim 32$ {\sc au} and $\sim 43$ {\sc au}.  Indeed, with our
choices of $\dot\theta$ we will find objects that are in orbits with $
a \in [30, 40]$\,{\sc au} and with moderate eccentricity, over most of
their orbital phase.
(Orbits beyond this range of semimajor axis comprise the remainder of
the 500 orbits in our survey. The results of that study will be
presented elsewhere.)
It is possible that some of the objects we found are
Centaurs near aphelion. However, since the density of Centaurs is
orders of magnitude smaller than the expected Kuiper belt density
(Levison \& Duncan 1995), it is unlikely we would find a Centaur in our
small field of view.

  Since we are observing on the ecliptic, every object we discover
must be passing through one of its nodes.  Thus the relationship
between inclination, $i$, and $\phi$ is simply $ i = \vert \phi -
\phi_E \vert$, where  $\phi_E = 0.3^\circ$ is a small correction due
to the shift from heliocentric to geocentric coordinates.
For every value of $i$ there are two values of
$\phi$, depending on which node the object is at.
We studied orbits with $\phi \in [0,30.3]$, which implies we
are sensitive to half the orbits with $i < 30^\circ$ (we have
not yet searched $\phi<0$ orbits).

On average, we found approximately 0.5 candidate objects per orbit
studied.  To demonstrate that we found real objects and to determine their
physical and dynamical characteristics, we combined the results of all
of our orbits with $i < 12^\circ$ to form a statistical sample.  Our
analysis found 53 candidate objects in the prograde orbit set,
but only 24 in the retrograde orbit set\footnote{These numbers differ
from those reported in IAUC 6163 because the present analysis is
restricted to objects with $V\leq28.1$; fainter objects were included in
the Circular.}.  A $\chi^2$ analysis shows that the probability that
these numbers are drawn from the same population is $9.5 \times
10^{-4}$.  There is no significant excess of candidates in our
prograde sample for $i>12^\circ$.

Figure~1 shows the magnitude distribution of candidate objects in the
prograde and retrograde sets of
orbits.  Notice the prograde candidates typically outnumber the
retrograde candidates in the bins brighter than our limiting magnitude,
and the brightest candidate discovered is in the prograde data set.
Visual inspection of several of the
brightest prograde candidates
shows that they are indeed detected in each frame and that
they appear stellar, confirming that they are very likely real.
We conclude that we are seeing excess candidates in the prograde orbit
set at $>$99.9\% ($1-9.5 \times 10^{-4}$) confidence level.  Since the
prograde and retrograde orbit sets were reduced in exactly the same way
and there should not be any systematic difference in the noise
characteristics, we conclude that we are seeing a population of
objects in the outer solar system that preferentially populates
prograde orbits.

The slope in the distribution shown in Figure~1 may appear surprisingly
steep.  However, the limiting magnitude of our survey is close to the
magnitude of the brightest object found in our field.  In such cases a
steep slope in the observed magitude distribution is inevitable,
independent of the intrinsic slope of the distribution.

We also compare the prograde and retrograde candidates
by looking at their inclination distributions.  In Figure~2,
we bin our results so that each plotted point contains a sum of the
data for all eccentricities at a given prograde inclination.
As a comparison, the dotted line
shows the mean of the number of candidates in the retrograde orbits
found in all the inclination bins (the individual retrograde orbit
values are shown as asterisks).  Thus, the dotted line can be viewed as
the zero point. Each prograde and retrograde bin includes eleven
different eccentricity values.  The figure shows a steady decrease in
the number of objects as a function of $i$.  The over abundance of
objects in the prograde orbits disappears for inclinations above
$\sim 15^\circ$.  Interestingly, we see a strong, statistically significant,
peak in the inclination distribution at about $3^\circ$.  Most of the
offset of the peak from 0$^\circ$ can be attributed to the
difference between  ecliptic and invariable plane inclinations and
the low probability of very small inclinations (which require orbit poles
in small areas of the celestial sphere; Marsden priv.~comm.).

The inclination distribution shown in Figure~2 adds evidence
that the excess objects detected in the prograde orbits are indeed
in the Kuiper belt.  As discussed in \S1, before the
discovery of 1992QB1 by Luu \& Jewitt in 1992, the primary argument
in support of the existence of the Kuiper belt was the theoretical
argument that it was required to be a flattened reservoir
in order to explain the low inclination distribution of the short-period comets
(Fern\'andez~1980; Quinn {\it et al.} 1990).
Our observations show that the
population that we are detecting is made up of low inclination
objects, consistent with models of the Kuiper belt and all
24 known 100--200\,{\sc km} objects in the belt.

Finally, it is possible to estimate the size distribution of the
objects we found.  To be conservative we only include objects brighter
than $28.6$.  The first step in determining this distribution is to
remove observational biases.  There are four such biases that must be
removed: (1) We first removed contamination due to false detections.
To accomplish this, we calculated the probability that a candidate was
real as a function of magnitude, $p_r(V)$, from the histograms shown in
Figure~1: $ p_r(V) = \left[N_{pro}(V) -
N_{retro}(V)\right]/N_{pro}(V)$.  In order to smooth out this function,
we fit a line to the resulting relationship. (2) We corrected for the
fact that our observations were only complete to $V=27.7$, but we were
seeing some objects down to $V \sim 29.0$.  We calculated the
probability that an object was discovered as a function of V magnitude,
$p_d(V)$.  We took $p_d(V) = 1$ for $V<27.7$ and assumed that it
linearly decreased as $V$ increased so that at $V = 28.6$, $p_d(V) =
0.5$.  A linear relationship is consistent with the results of the
tests we performed using artificial objects, as discussed in \S2.
(3) We corrected for the fact that large objects could be seen
at greater distances from the Sun than small objects by calculating
the fraction, $f_V$, of the volume of
space that we searched (as defined by our choice of orbits) in which an
object of a particular radius would actually have been visible to us.
The inner edge of the volume was taken to be $25.5$\,{\sc au},
corresponding to the perihelion distance of an object in Neptune's 2:3
mean motion resonances with $e=0.3$.  The outer edge of the volume was
taken to be $35.5$\,{\sc au}, corresponding to the perihelion distance
of an object with the same semi-major axis but with $e=0.1$.  We
performed this calculation assuming a limiting magnitude of $V=28.6$.
And (4) we corrected for the fact that the effective size of our field
is a function of $\dot\theta$ and $\phi$.  This can be seen by recognizing
that the larger the drift rate, the more we had to
shift the frames to compensate for the orbits of the body, thus the
smaller the effective field.  We determined by visual inspection
the fraction of the field, $f_F$, over which we could find objects.
For each of our prograde candidates, we calculated its current
heliocentric distance, $r$, under the assumption that it was in
Neptune's 2:3 mean motion resonance and was at perihelion.  We then
calculated its radius, $R$, assuming an albedo of $0.04$.  We assigned
a weight to the candidate, $ w = p_r / (p_d\, f_V\, f_F)$.  This weight
is an estimate of this object's contribution to the size distribution
of real Kuiper belt objects.

   The solid curve in Figure~3 shows the size distribution of our
candidate objects as derived using the above procedure.  Note that
under the assumption of an albedo of 4\%, we found objects as small as
$5$\,{\sc km} in radius!  Also shown in Figure~3 are two hypothetical
size distributions of the form $n(R) dR \propto R^{-b}\, dR$.  A
distribution with $b=3$ is believed to be a reasonable representation
of the size distribution of the known short-period comets (Shoemaker \&
Wolfe~1982).  A distribution with $b\sim 5$ is required for larger
objects in the Kuiper belt in order to reconcile the number of comets
believed to be there and the number of known 100--200\,{\sc km} objects
(Duncan {\it et al.} 1995).  Although there are significant uncertainties in
our derived size distribution due to the small number of objects
detected, as well as to inaccuracies in determining the magnitudes of
the objects and their exact heliocentric distances, it appears there is
a reasonable size distribution for our objects.

\section{Conclusion}

We have used HST's WFPC2 to deduce that there are a substantial number
of Halley-sized objects in a disk-like structure in the region near and
beyond the orbit of Neptune.  We found 29 objects with
radii ranging from 5 to 10\,{\sc km} (4\% albedo) in the
orbits discussed in \S3.  Our observations imply there are $\sim
25,000$ objects degree$^{-2}$ brighter than $V=28.6$.  Thus, there
are $>2 \times 10^8$ comets in this size range in orbits similar to the
ones we have studied here.  Combining this number with the size
distribution shown in Figure~3 and assuming a density of $0.5$ {\sc gm
cm$^{-3}$} implies a total of $0.02$ Earth masses of 5--10\,{\sc km}
comets in {\it this region} of the Kuiper belt.  We note that the
derived radii are uncertain to a factor of a few  and the
density of comets is highly uncertain.  Therefore, it is possible that
the mass in this region of the Kuiper belt differs from that quoted
here by as much as an order of magnitude.

The observations presented here represent the first reported detections of
objects the size of typical short-period comets {\it in their native
reservoir}.  Their detection shows that the idea
is indeed correct that a significant
number of planetesimals extends the solar system past the region of the
planets.  These objects appear to be confined to a
moderate-thickness disk near the plane of the ecliptic, as is necessary
if these objects are the source of most of the short-period comets.
For the first
time, one can point to a definitive region of the solar nebula for the
origin of short-period comets.

\acknowledgments

We thank A.~Whipple, W.~Cochran, J.~Bally, B.~Marsden, S.~Tremaine, and
P.~Weissman for helpful discussions.  Support for this work was
provided by NASA through GO-5312.01-93A from the Space Telescope
Science Institute which is operated by the Association of Universities
for Research in Astronomy, Inc., under NASA NAS5-26555.

\newpage
\section{Figure Captions}

\noindent {\bf Figure~1 --- }  A histogram of the number of Kuiper belt
candidates we found as a function of V magnitude.  The solid green and dotted
red curves refer to candidates discovered in our prograde and retrograde
orbit sets, respectively.  The dashed line shows our ``limiting''
magnitude ($i.e.$ at which the detection probability
is 0.5) at $V=28.6$.

\medskip

\noindent {\bf Figure~2 --- }  The solid dots show the number of candidate
objects found in our prograde orbit set as a function of their orbital
inclination; the asterisks show the candidates from our control
sample.  The dotted horizontal line shows the mean number of candidates
found in the retrograde frames and can be interpreted as the number of
false detections per inclination bin.  Note the significant
number of excess candidates at low inclination, but the excess becomes
insignificant for $i \gapprox 15^\circ$.

\medskip

\noindent {\bf Figure~3 --- } The size distribution of our excess
candidate objects is shown in the form of a histogram (solid curve) of
the number of objects as a function of radii, $R$.  Also shown are two
hypothetical differential size distributions
of the form $n(R) dR \propto R^{-b}\, dR$ with $b=3$
(dotted curve) and $b=5$ (dashed curve), respectively.  The total area
under the each curve is normalized to 1.

\epsfxsize=7.25in \epsfbox{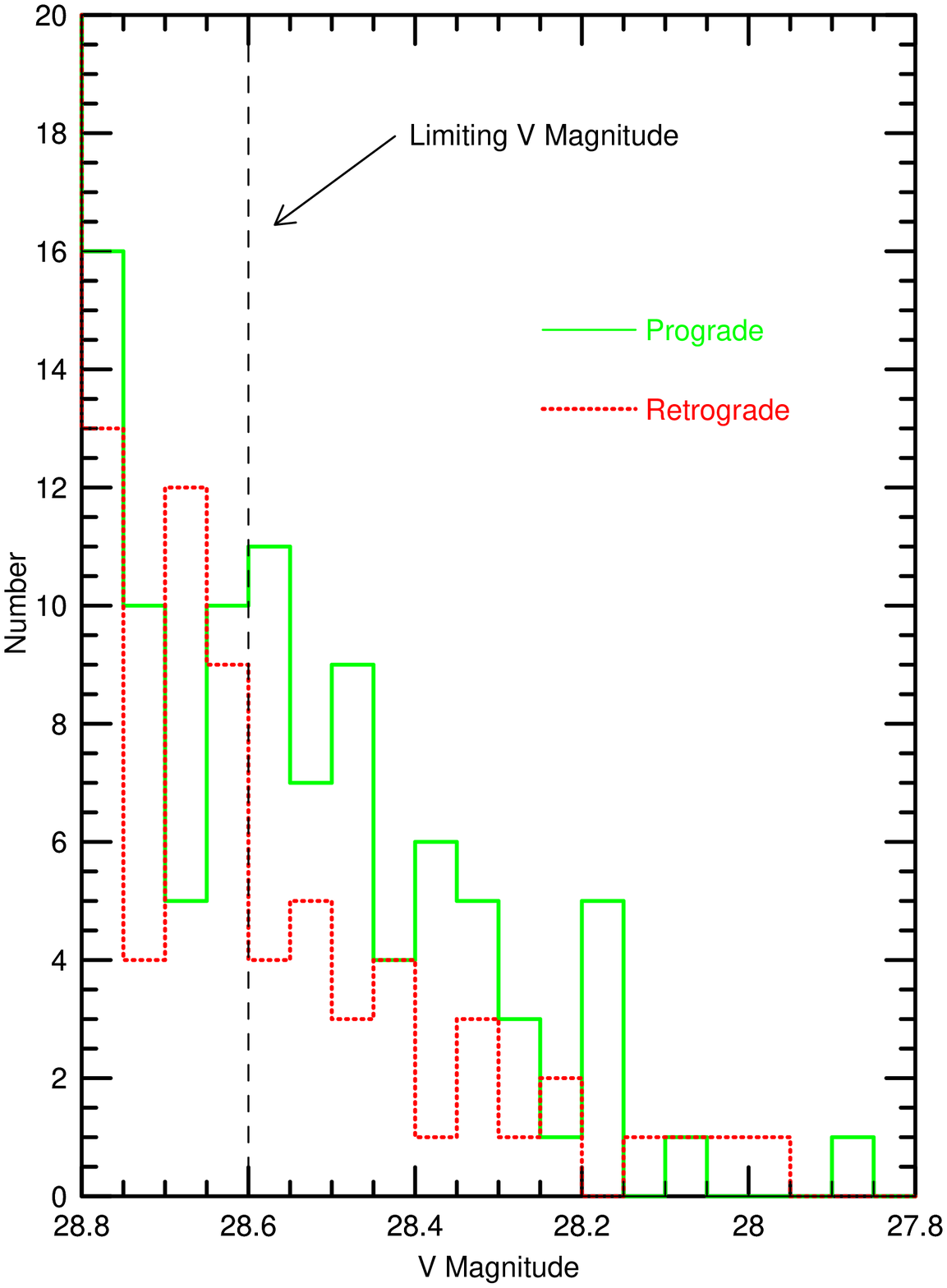}
\epsfxsize=7.25in \epsfbox{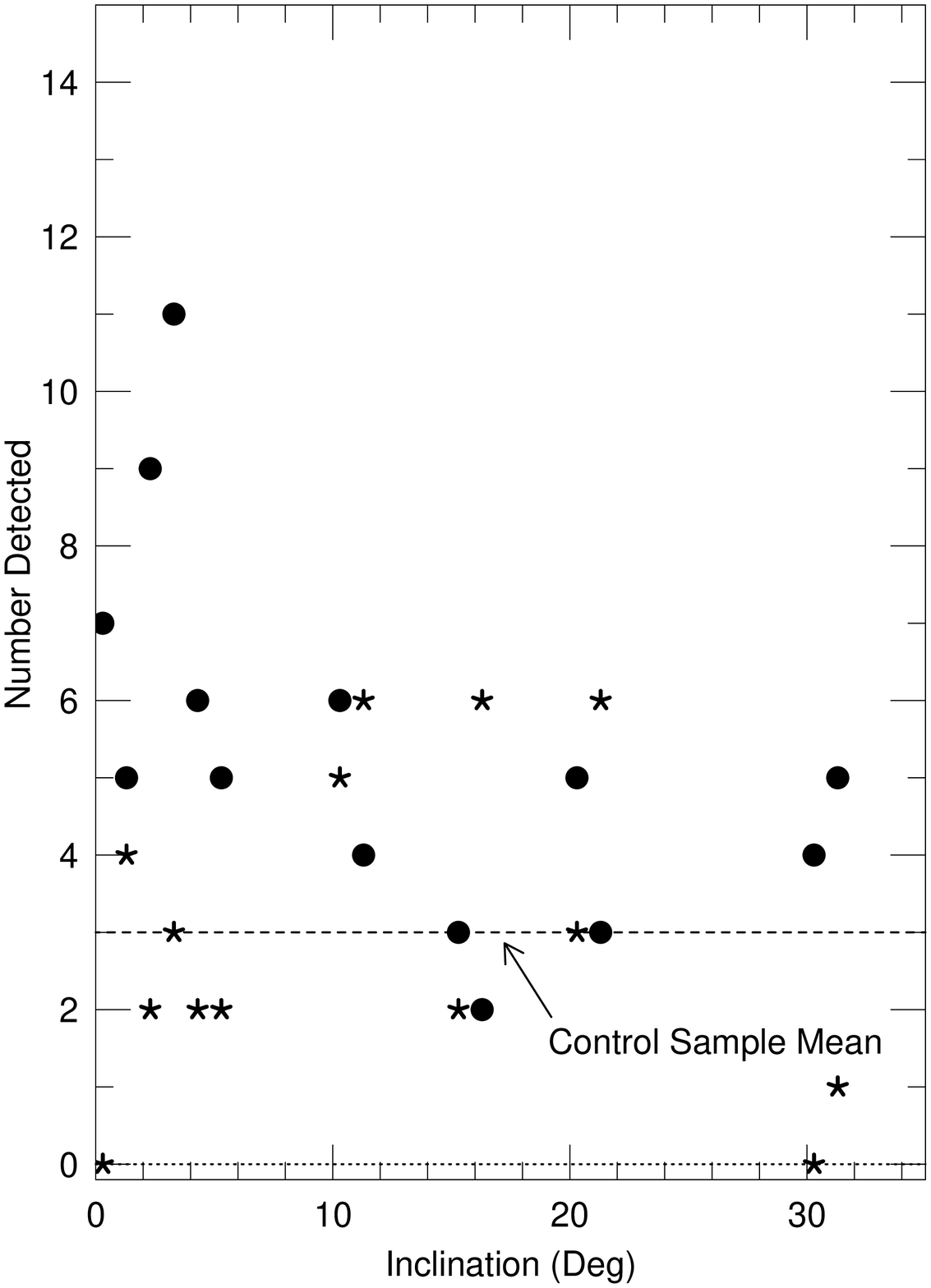}
\epsfxsize=7.25in \epsfbox{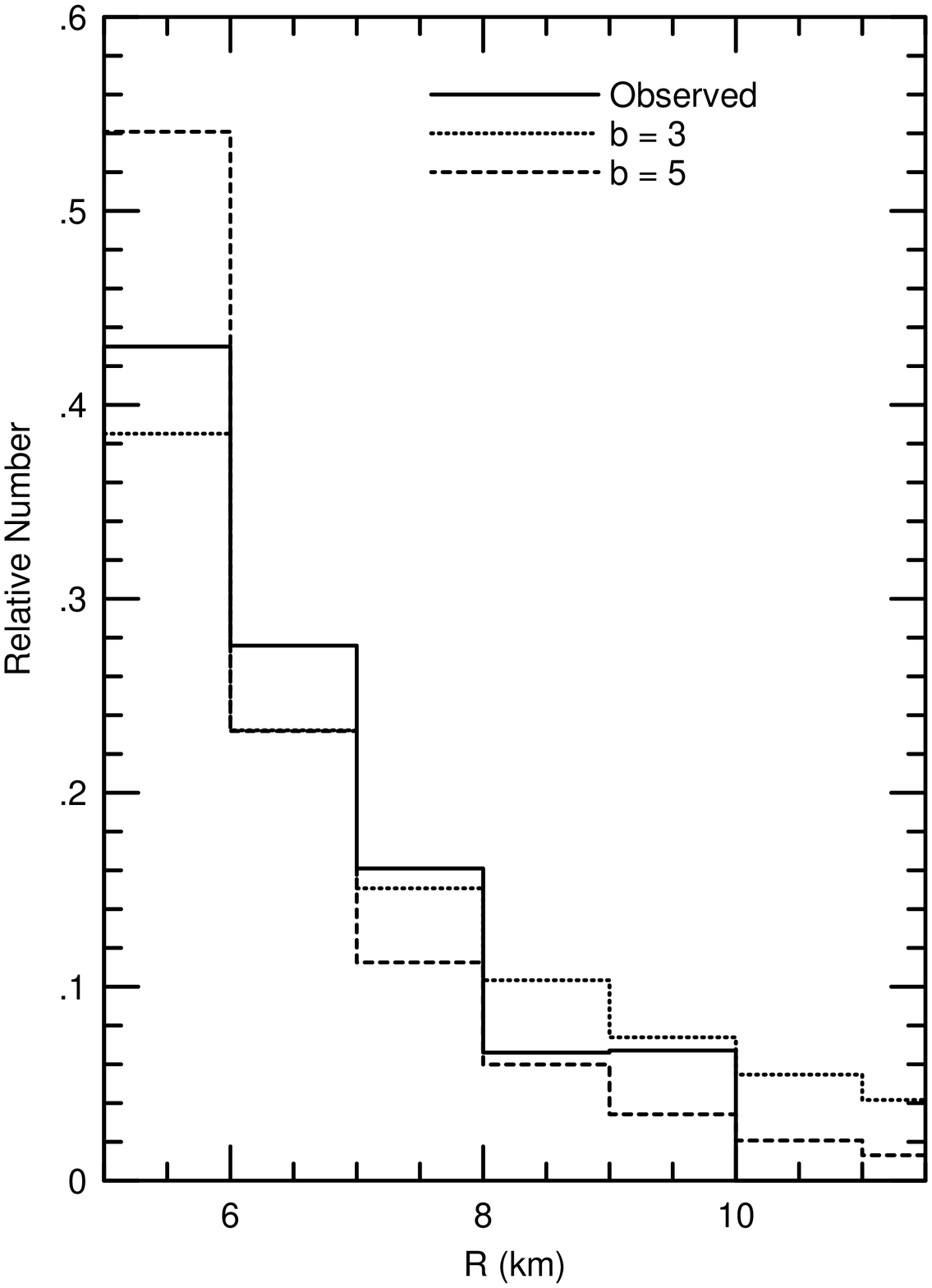}


\begin{references}
\reference Duncan, M, Levison, H., \& Budd, M. 1995. \aj, in press
\reference Duncan, M., Quinn, T., \& Tremaine, S. 1988, \apj, 328, L69
\reference Edgeworth, K.E. 1949. \mnras, 109, 600
\reference Everhart, E. 1972,  \apj, 10, L131
\reference Everhart, E. 1977, in {\it Comets--Asteroids--Meteorites,}
     ed. A.H. Delsemme, (Ohio: University of Toledo Press), 99
\reference Fern\'andez, J. A. 1980, \mnras, 192, 481
\reference Hainaut, O., West, R. M., Smette, A. \& Marsden, B. G.
     1994. A\&A 289,311
\reference Holman, M., \& Wisdom, J. 1993. \aj, 105, 1987
\reference Jewitt D. \& Luu, J. 1992.   IAU Circulars, 5611
\reference Jewitt D., \& Luu, J.~1995. \aj, 109, 1867.
\reference Kuiper, G. P. 1951. In {\it Astrophysics: A Topical Symposium},
ed. J.A. Hynek, (New York: McGraw Hill), 357
\reference Levison, H. \& Duncan, M. 1993. \apjl, 406, L35
\reference Levison, H. \& Duncan, M. 1995. In preparation
\reference Marsden, B. G. 1994. IAU Circulars 5983
\reference Marsden, B. G. 1995. MPEC 1995-C17
\reference Morbidelli, A., Thomas, F. \& Moons, M.~1995. Icarus
    submitted
\reference Newton, H. 1893, {\it Mem. Natl. Acad. Sci.}, 6, 7
\reference Quinn, T., Tremaine, S., \& Duncan (1990). \apj, 355, 667.
\reference Shoemaker, E., \& Wolfe, R. 1982  in {\it Satellites of Jupiter}
(ed. Morrison, D.), 277.
\reference Stern, S.A. 1995, To appear in `Completing the Inventory of the
Solar System', ed. E. Bowell. (ASP Conference Series)
\reference Weissman, P. 1995. {\it Ann. Rev. Astron. Astrophy.} 33, 327
\end{references}
\end{document}